\documentclass[aps,preprint,floats,superscriptaddress,nofootinbib]{revtex4}
\usepackage{amsmath,amssymb}
\usepackage{url}
\usepackage{graphicx}

\newlength{\defbaselineskip}
\newcommand{\setlinespacing}[1]%
           {\setlength{\baselineskip}{#1 \defbaselineskip}}

\begin{document}

\preprint{
\hfill
\begin{minipage}[t]{3in}
\begin{flushright}
\vspace{0.0in}
FERMILAB--PUB--07--666--A\\
\end{flushright}
\end{minipage}
}

\hfill$\vcenter{\hbox{}}$

\vskip 0.5cm

\title {Neutralino Dark Matter and Trilepton Searches in the MSSM}
\author{Dan Hooper}
\affiliation{Theoretical Astrophysics Group, Fermilab, Batavia, USA}

\author{Tilman Plehn}
\affiliation{SUPA, School of Physics, University of Edinburgh, Scotland}

\author{Alberto Vallinotto}
\affiliation{Institut d'Astrophysique de Paris, CNRS-UMR 7095, 
             Universit\'{e} Paris VI Pierre et Marie Curie, Paris, France}

\date{\today}

\bigskip

\begin{abstract}
    Searches for supersymmetry are among the most exciting physics goals
  at Run~II of the Tevatron. In particular, in supersymmetric models
  with light charginos, neutralinos and sleptons, associated
  chargino--neutralino production can potentially be observed as
  multi--lepton events with missing energy. We discuss how, in the
  generic TeV--scale MSSM, the prospects for these chargino--neutralino
  searches are impacted by cosmological considerations, namely the
  neutralino relic abundance and direct detection limits. We also
  discuss what an observation of chargino--neutralino production at
  the Tevatron would imply for the prospects of future direct dark
  matter searches without assuming specific patterns of supersymmetry breaking.
\end{abstract}

\maketitle

\newpage

\section{Introduction}

A consensus has formed within the astrophysics community in support of
the conclusion that the majority of our universe's mass takes the form
of cold, collisionless dark matter~\cite{dmreview}. Despite the
very large body of evidence in favor of dark matter's existence, the
nature of this elusive substance remains unknown. Of the many dark
matter candidates to have been proposed, one of the most compelling
and most often studied is the lightest neutralino in $R$-parity
conserving models of supersymmetry~\cite{jungman}.\bigskip

Among the most prominent missions of the Tevatron's Run~II are its
searches for supersymmetry. Results from Tevatron searches for squarks
and gluinos~\cite{tevsquarksgluinos}, neutralinos and
charginos~\cite{tevneucha}, stops and sbottoms~\cite{stopsbottom}, and
the Higgs bosons of the Minimal Supersymmetric Standard Model
(MSSM)~\cite{tevhiggs} have each recently been published. While no
evidence for supersymmetry has yet been found, in many cases these
results represent the strongest limits to date. Although the Tevatron
is not well suited to directly place limits on the properties of the
lightest neutralino, the results of these other searches can have
considerable implications for the nature of this dark matter agent.

One of the prime channels for observing supersymmetry at the Tevatron
is associated neutralino--chargino
production~\cite{trilep_early,trilep_newer}.  These particles can
decay to the lightest neutralino and leptons through the exchange of
either sleptons or gauge bosons, resulting in events featuring three
leptons and missing energy. The results of these searches are somewhat
model dependent, but the current results from CDF and D0 can be used
to exclude charginos as heavy as approximately 150~GeV in some models,
well beyond LEP's chargino mass limit of 104~GeV. By the end of Run
II, the Tevatron is expected to exclude selected models with charginos
not far below 200~GeV.\bigskip

Neutralino dark matter can be detected through its elastic scattering
with nuclei. Experimental efforts designed to observe such events are
known as direct detection. The prospects for this class of techniques
depends on the composition of the lightest neutralino, as well as on
the masses and couplings of the exchanged squarks and Higgs bosons.
Generally speaking, information from collider searches for
supersymmetry, whether detections or constraints, can be used to
better estimate the prospects for the detection of neutralino dark
matter. The relationship between Tevatron and LHC searches for heavy
MSSM Higgs bosons and direct searches for neutralino dark matter has
been studied in detail elsewhere~\cite{marcela}. Here, we return to
this theme, but focussing on searches for trilepton events from
associated neutralino--chargino production at the Tevatron (see also Ref.~\cite{trilep_dm}).

In the past, it has been possible to link the Tevatron's trilepton
signature to other signatures for new physics, for example the decay
$B_s \to \mu \mu$~\cite{bmumu}. Such links rely for example on a
correlation between light sleptons and small values of $\tan\beta$ for
the chargino and neutralino decays on one hand and the pseudoscalar
Higgs boson mass and large values of $\tan\beta$ in flavor physics on
the other. Naively, similar correlations should be present when the
dark matter candidates annihilate mainly through an $s$-channel Higgs
boson and the trilepton signature requires relatively light
supersymmetric scalars. Moreover, one could imagine correlations
between these signals in the co-annihilation region, if the lightest
slepton is mass degenerate with the lightest neutralino, limiting the
visibility of the trilepton channel. However, these fairly obvious
correlations rely on a series of assumptions. First, the different
MSSM scalar masses have to be correlated. Secondly, the light scalar
masses should in some way be linked to the lightest neutralino mass
and to the mass difference between the light chargino and neutralino.
Last but not least, the dark matter particle should annihilate
dominantly through one channel. The aim of this analysis is to determine
how much of a correlation between dark matter and Tevatron searches
survives if we assume only a TeV--scale MSSM spectrum with no specific patterns of supersymmetry breaking.\bigskip

This article is structured as follows: In Section~\ref{sec:tri}, we
discuss the searches for associated chargino-neutralino production at
the Tevatron. In Section~\ref{sec:dm}, we turn our attention to the
thermal relic abundance of neutralinos, focussing on those models
within the reach of the Tevatron and the correlation between Tevatron
measurements and the neutralino's relic density. In
Section~\ref{sec:direct}, we discuss direct detection prospects for
such models and the correlations between those and Tevatron
observations.  Finally, we summarize our results and conclusions in
Section~\ref{sec:conclusion}.

\section{Neutralino--Chargino Searches at the Tevatron}
\label{sec:tri}

In many supersymmetric models, associated chargino-neutralino
production can occur with a cross section on the order of a picobarn
at Run~II of the Tevatron (1.96~TeV center-of-mass collisions). These
particles can each subsequently decay to the lightest neutralino and
leptons ($\chi^{\pm}_1 \rightarrow \chi^0_1 \, l^{\pm} \nu$, $\chi^0_2
\rightarrow \chi^0_1 \, l^{\pm} l^{\mp}$), either through the exchange
of charged sleptons or gauge bosons. This can lead to distinctive
trilepton plus missing energy events which, in some supersymmetric
models, could be identified over Standard Model backgrounds.

\begin{figure}[t]
\includegraphics[width=3.3in,angle=0]{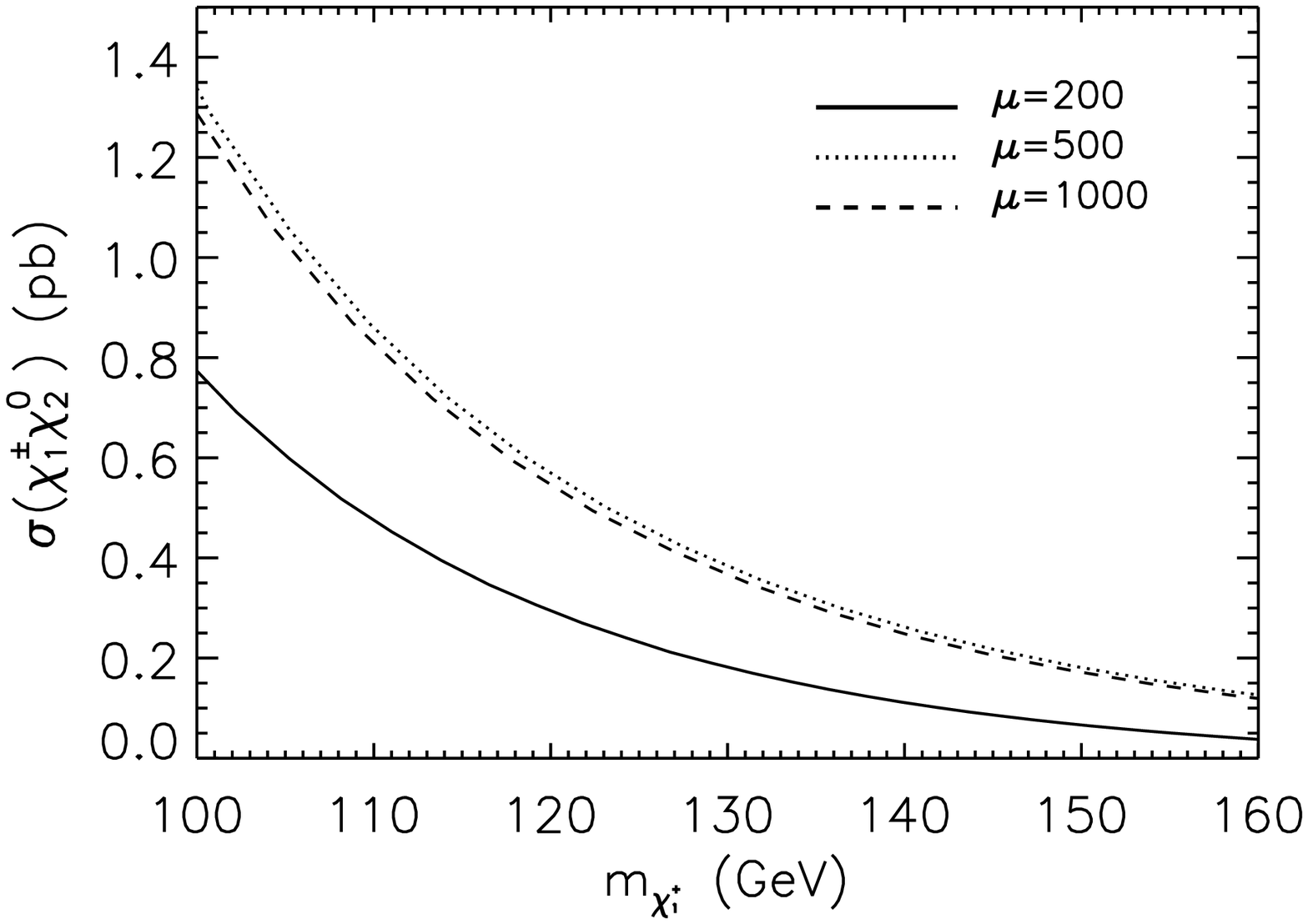}
\includegraphics[width=3.3in,angle=0]{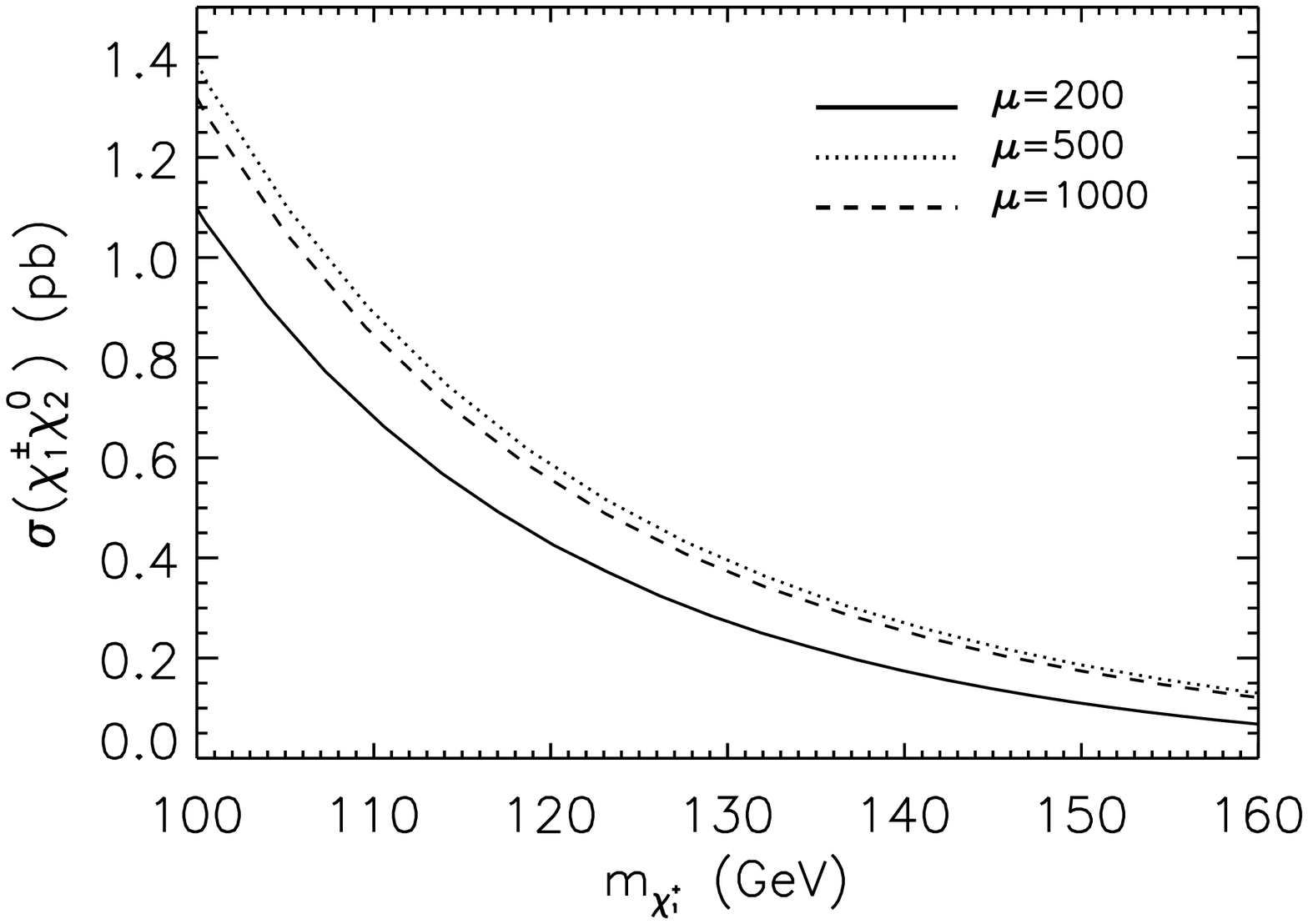}
\\
\caption{The cross section for associated chargino-neutralino 
  production at Tevatron Run~II, as a function of the lightest
  chargino mass, for various choices of $\tan \beta$ (3 and 60, in the
  left and right frames, respectively) and $\mu$~\cite{prospino}.  In
  these Figures we use $2 M_1=M_2$,
  $m_{\tilde{q}}=m_{\tilde{l_L}}=m_A=$500~GeV and
  $A_t=A_b=A_{\tau}=$0.}
\label{fig:sig}
\end{figure}

In order for SUSY--trilepton events to be extracted at the Tevatron,
however, the underlying supersymmetric model must possess a number of
rather specific features. In particular, the $\chi^{\pm}_1$ and
$\chi^0_2$ must both be light. In Fig.~\ref{fig:sig} we plot the
associated chargino-neutralino production cross section as a function
of the lightest chargino mass for various values of $\tan \beta$ and
$\mu$. The cross section drops rapidly for heavy chargino/neutralino
masses. Additionally, in order to be identified at the Tevatron,
$\chi^{\pm}_1$ and $\chi^0_2$ decays must each occur with large
branching fractions to charged leptons, which means that the
supersymmetric mass spectrum is arranged such that chargino and
neutralino decay primarily to charged sleptons rather than to (off-shell)
gauge bosons or squarks, each of which lead to significant branching
fractions to jets. Furthermore, $WZ$ production leads to a Standard
Model background of trileptons plus missing energy
from which any SUSY-trileptons must be separated. To accomplish this,
the analyses of CDF and D0 each include kinematic cuts on observables
like $m_{\ell\ell}$, designed to remove backgrounds. They reduce the
efficiency for supersymmetric events with charginos and/or neutralinos
decaying through gauge bosons essentially to zero.\bigskip

\begin{figure}[t]
  \includegraphics[width=3.3in,angle=0]{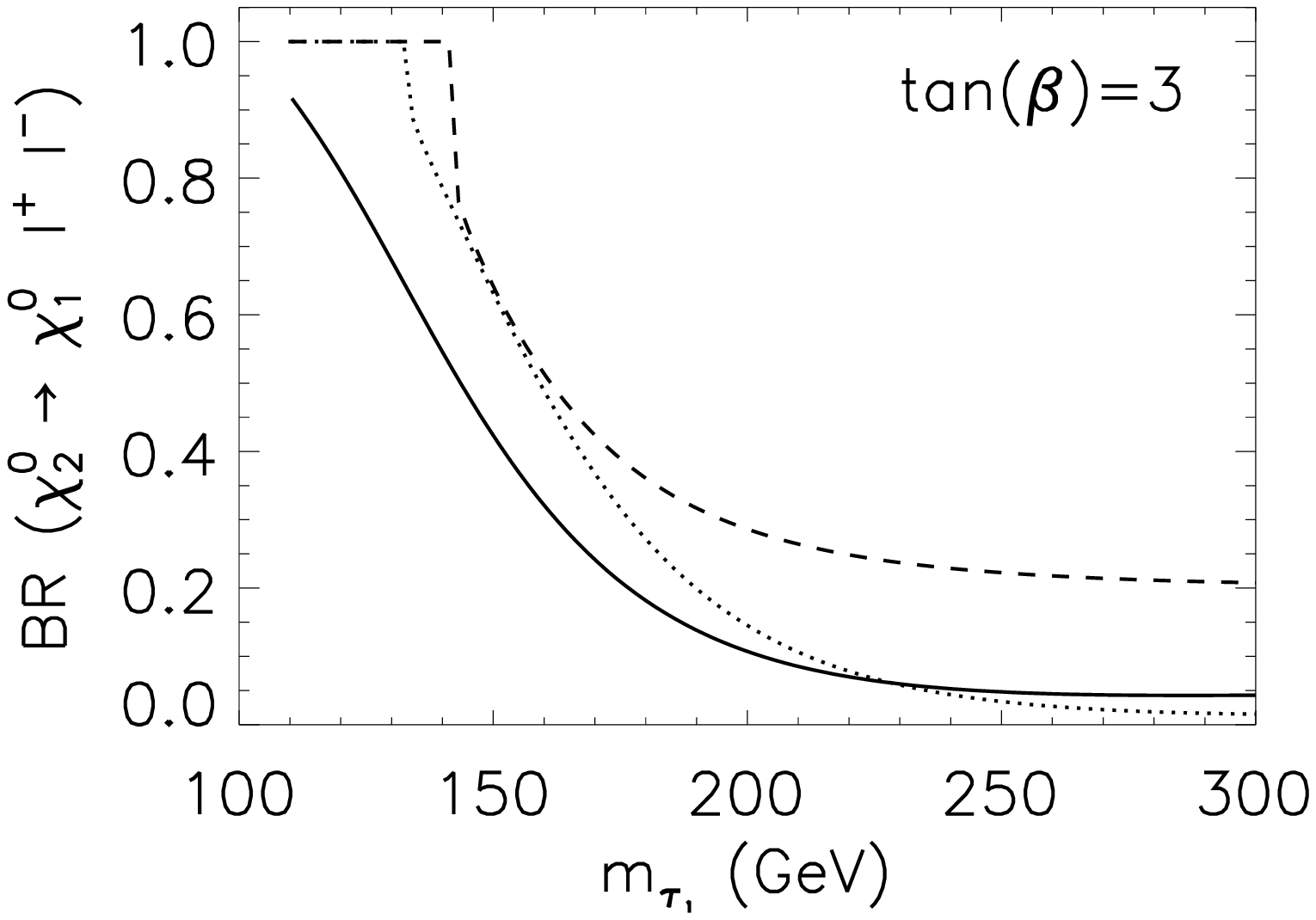}
  \includegraphics[width=3.3in,angle=0]{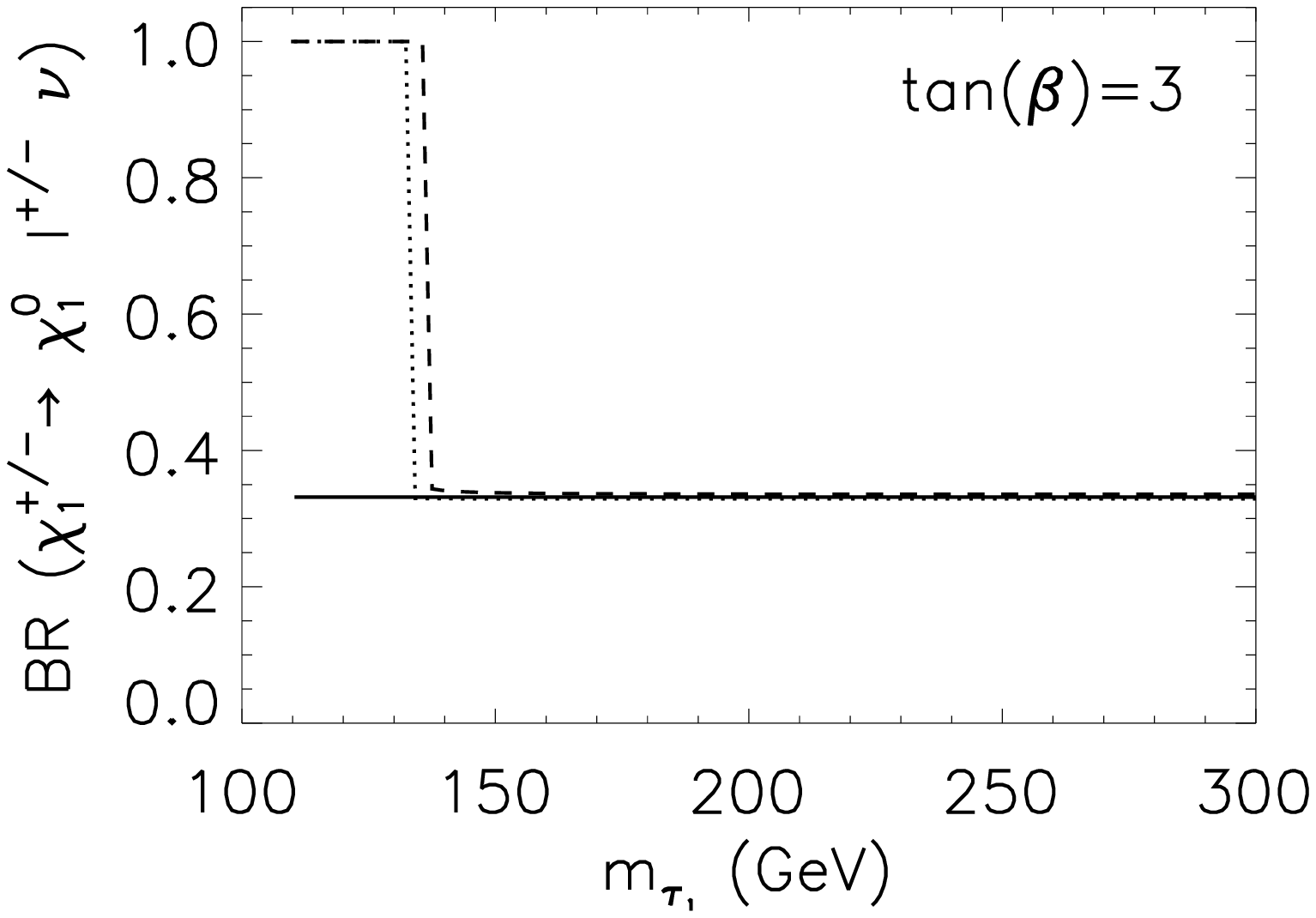}
  \\
  \includegraphics[width=3.3in,angle=0]{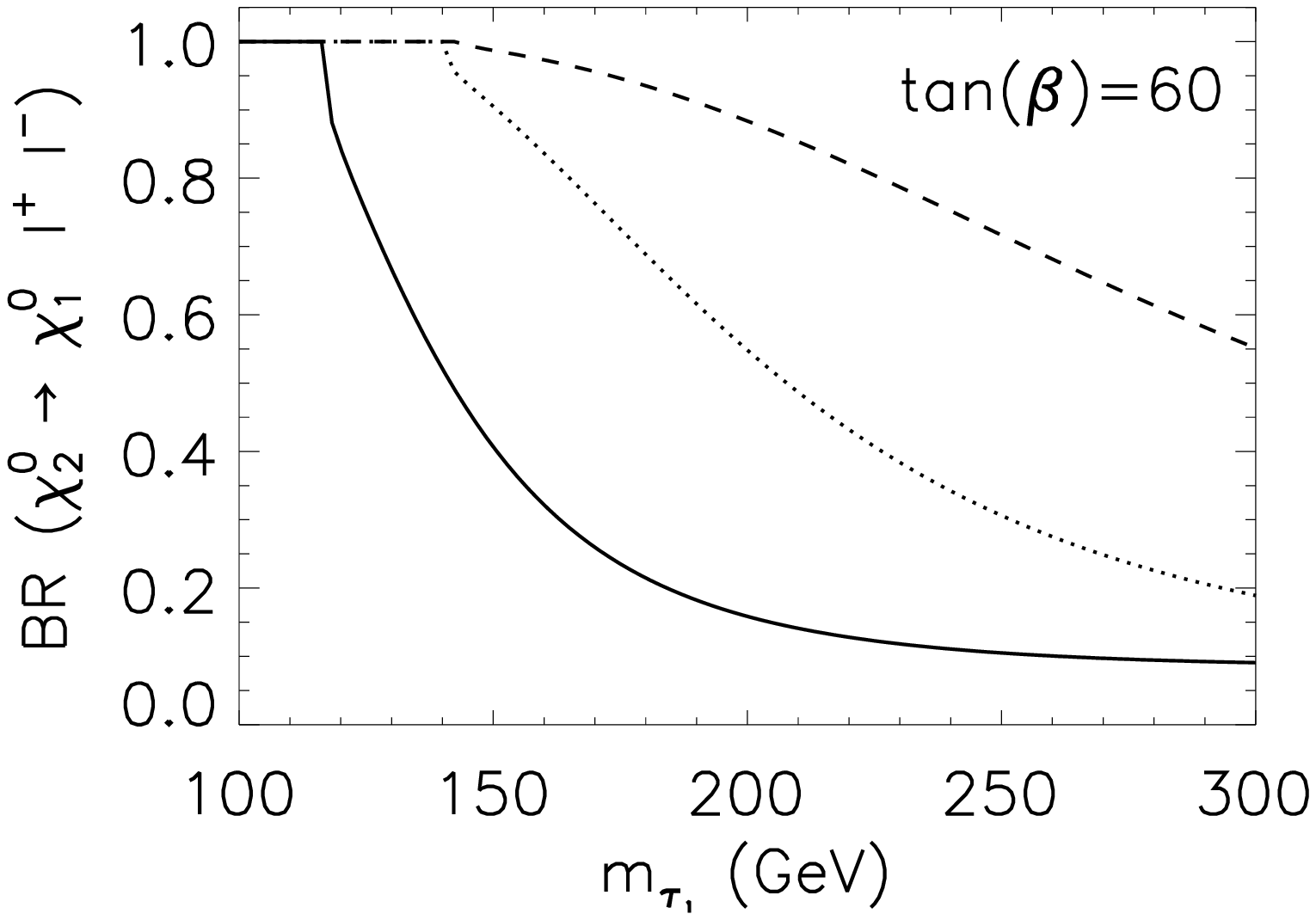}
  \includegraphics[width=3.3in,angle=0]{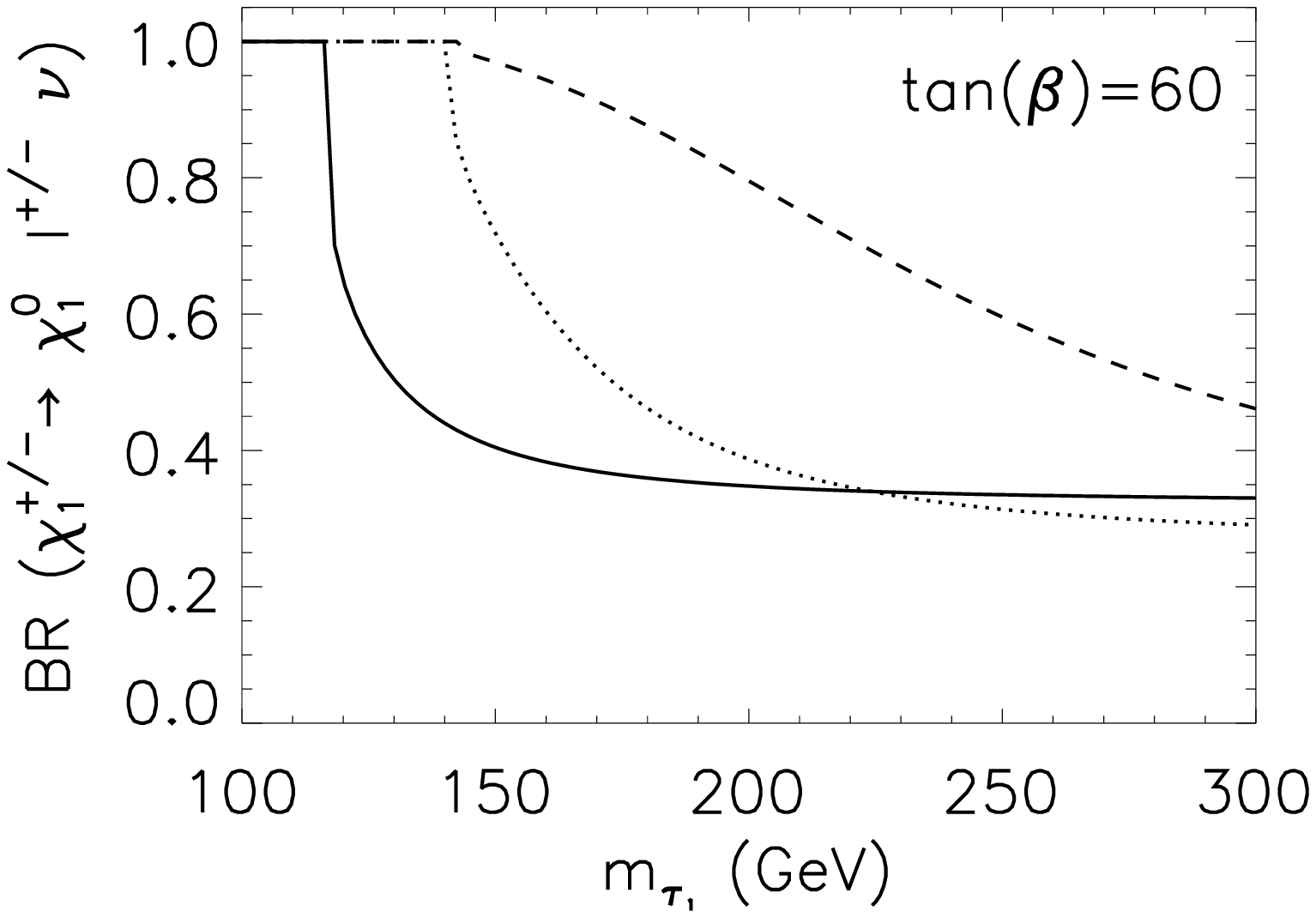}
  \\
\caption{The branching fractions of $\chi^0_2$ and $\chi^{\pm}_1$ decays 
  to final states with charged leptons, as a function of the lightest
  stau mass and for $\tan \beta= 3, 60$. The three lines denote
  $\mu=$200~GeV (dashed), 500~GeV (dotted) and 1~TeV (solid). We have
  also used $2 M_1=M_2=$140~GeV,
  $m_{\tilde{q}}=m_{\tilde{l_L}}=m_A=$500~GeV and
  $A_t=A_b=A_{\tau}=$0. In order for the combination of $\chi^0_2
  \chi^{\pm}_1$ to decay mostly to final states with three charged
  leptons, the lightest stau (possibly along with other charged
  sleptons) must be rather light.}
\label{fig:BR}
\end{figure}

To ensure large branching fractions for charginos and neutralinos
through slepton exchange, the lighter sleptons must be quite light. To
avoid large neutralino or chargino branching fractions to neutrinos,
the sneutrino masses (along with the left handed charged sleptons)
must be somewhat heavier. Unless we want to break the $SU(2)$ symmetry
between charged slepton and sneutrino masses, this means the lightest
charged slepton should be dominantly right handed, independent of
unification assumptions. In Fig.~\ref{fig:BR}, we plot the branching
fractions of charginos and neutralinos to trileptons (including
electrons, muons and taus), as a function of the lightest stau mass,
for various choices of $\mu$ and $\tan \beta$.

Limits from the CDF and D0 collaborations have been placed on the
combined cross section for associated neutralino-chargino production
and branching fractions to three leptons. In particular, D0 has
published results for their search for events with three leptons (at
least two of which are electrons or muons) plus missing energy using
the first 320 pb$^{-1}$ of data from Run~II~\cite{Abazov:2005ku}. They
find a rate consistent with the predictions of the Standard Model, and
use this to place constraints on supersymmetry. CDF has published the
results of their search for events with two like-sign leptons
(electrons or muons) and missing energy using 1 fb$^{-1}$ of data from
Run~II~\cite{Abulencia:2007rd}. In this analysis, 13 events were
observed, a slight excess compared to the 7.8 predicted by the
Standard Model (corresponding to a chance probability of 7\%). More
recently, CDF has published the results of their combined search for
associated neutralino-chargino production. These findings are
consistent with Standard Model expectations~\cite{:2007nz}. In
addition to these published results, a number of preliminary results
from CDF~\cite{:2007mu} and D0~\cite{d0preliminary} searches for
trilepton plus missing energy events have been reported.

\begin{figure}[t]
\begin{center}
\includegraphics[width=3.3in,angle=0]{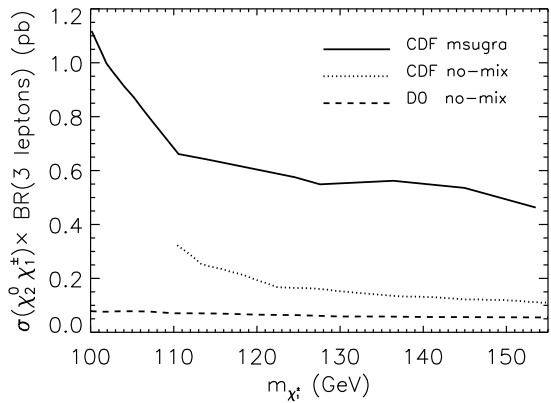}
\end{center}
\caption{The current limits on associated neutralino-chargino production 
  from CDF and D0 searches for SUSY-trilepton events. See text for
  details.}
\label{fig:limits}
\end{figure}

In Fig.~\ref{fig:limits}, we show the current limits from CDF and D0
in this channel. The limits from CDF are shown for two scenarios,
labelled ``mSUGRA'' and ``no-mixing''. In the mSUGRA scenario, the
masses of the staus are determined within the context of the mSUGRA
model, which leads to the lightest stau being considerably less
massive than the other sleptons and, in turn, to large branching
fractions for chargino and neutralino decays to taus. In the no-mixing
scenario, decays to taus, muons and electrons are approximately
equally common. As taus are more difficult to identify than other
leptons, the CDF limit in the mSUGRA scenario is considerably weaker
than in the no-mixing case. Also shown is the D0 limit for the
no-mixing scenario. It is more stringent than the limit from CDF, in
part, because D0's result is slightly stronger than expected.  By the
end of Run~II, the limits from each of these experiments are expected
to improve by a factor of approximately 5 to 10.

\section{Thermal Abundance of Neutralinos within Tevatron Reach}
\label{sec:dm}

In $R$-parity conserving models in which the lightest neutralino is
the lightest supersymmetric particle (LSP), such particles fall out of
thermal equilibrium when the rate of Hubble expansion begins to
dominate over their annihilation rate. The resulting density of
neutralino dark matter in the universe today is related to its
annihilation cross section:
\begin{equation}
 \Omega_{\chi^0_1} h^2 \approx 
  \frac{1.04 \times 10^9 \, x_F}
       {M_{\rm Pl} \, \sqrt{g^\star} \, \langle \sigma v \rangle},
\end{equation}
where $\langle \sigma v \rangle$ is the thermally averaged
neutralino--neutralino annihilation cross section, $g^\star$ is the
number of relativistic degrees of freedom available at the temperature
of freeze-out, and $x_F \equiv m_{\chi^0_1}/T_F$, where $T_F$ is the
temperature of freeze-out. For neutralinos (and other species of
electroweak scale WIMPs), $x_F$ falls in the range of 20-30. The
thermally averaged annihilation cross section can be written as
$\langle \sigma v \rangle \approx a + 3b/x_F$, where $a$ and $b$ are
terms in the expansion $\sigma v = a + b v^2 +
\vartheta(v^4)$.\bigskip

The neutralino annihilation cross section depends on the details of
the supersymmetric model, including the composition of the LSP and the
masses and mixings of the exchanged sparticles and Higgs bosons. The
four neutralinos of the MSSM are mixtures of the superpartners of the
photon, $Z$ and neutral Higgs bosons. 
The neutralino mass matrix is
diagonalized into mass eigenstates by a unitary rotation $N^*
M_{\chi^0} N^{-1}$.  Hence, we can describe the lightest neutralino as
a mixture of gauginos and higgsinos:
\begin{equation}
\chi^0_1 = N_{11}\tilde{B}     +N_{12} \tilde{W}^3
          +N_{13}\tilde{H}_1 +N_{14} \tilde{H}_2.
\end{equation}

Although no accelerator bounds have been placed on the mass of the
lightest neutralino directly~\cite{how_light}, LEP II has placed a
lower limit of 104~GeV on the mass of the lightest chargino, which is
in turn related (at tree level) to $M_2$, $\tan \beta$ and $\mu$,
\begin{equation}
m_{\chi^{\pm}_{1}} = \frac{1}{\sqrt{2}} 
                     \left[ |M_2|^2+|\mu|^2+2 m^2_W 
                           - \sqrt{(|M_2|^2+|\mu|^2+2 m^2_W)^2 
                                    - 4|\mu M_2 - m^2_W \sin 2 \beta|^2} 
                     \right]^{1/2}.
\label{eq:charginomass}
\end{equation}
The LEP II bound, therefore, leads to a constraint of $|M_2|, |\mu| >
104$~GeV. Since we are interested in the case in which the
$\chi^{\pm}_1$ and $\chi^0_2$ are within the reach of the Tevatron,
and yet significantly heavier than the lightest neutralino, we are
forced to consider values of $M_1$ smaller than $M_2$ and $|\mu|$. If
$M_1$ is considerably smaller than $M_2$ and $|\mu|$, the lightest
neutralino will be largely bino-like, with a small higgsino admixture:
\begin{equation}
|N_{11}| \sim 1 \; , 
\qquad  
\frac{|N_{13}|^2}{|N_{11}|^2} \approx 
 \frac{m^2_Z \, \sin^2 \theta_W \, \sin^2 \beta}
      {|\mu|^2} \sim 0.01 \, \bigg(\frac{200 \, \rm{GeV}}{|\mu|}\bigg)^2 \; ,
\qquad 
|N_{14}|^2 < |N_{13}|^2 \; . 
\label{eq:n13}
\end{equation}
The mass of the lightest neutralino in this scenario is approximately given by
\begin{equation}
 m_{\chi^0_1} \approx 
 M_1 - \frac{m^2_Z \sin^2 \theta_W (M_1 + \mu \sin 2\beta)}{\mu^2 - M_1^2}. 
\label{eq:lspmass}
\end{equation}
%

\begin{figure}[t]
\includegraphics[width=3.1in,angle=0]{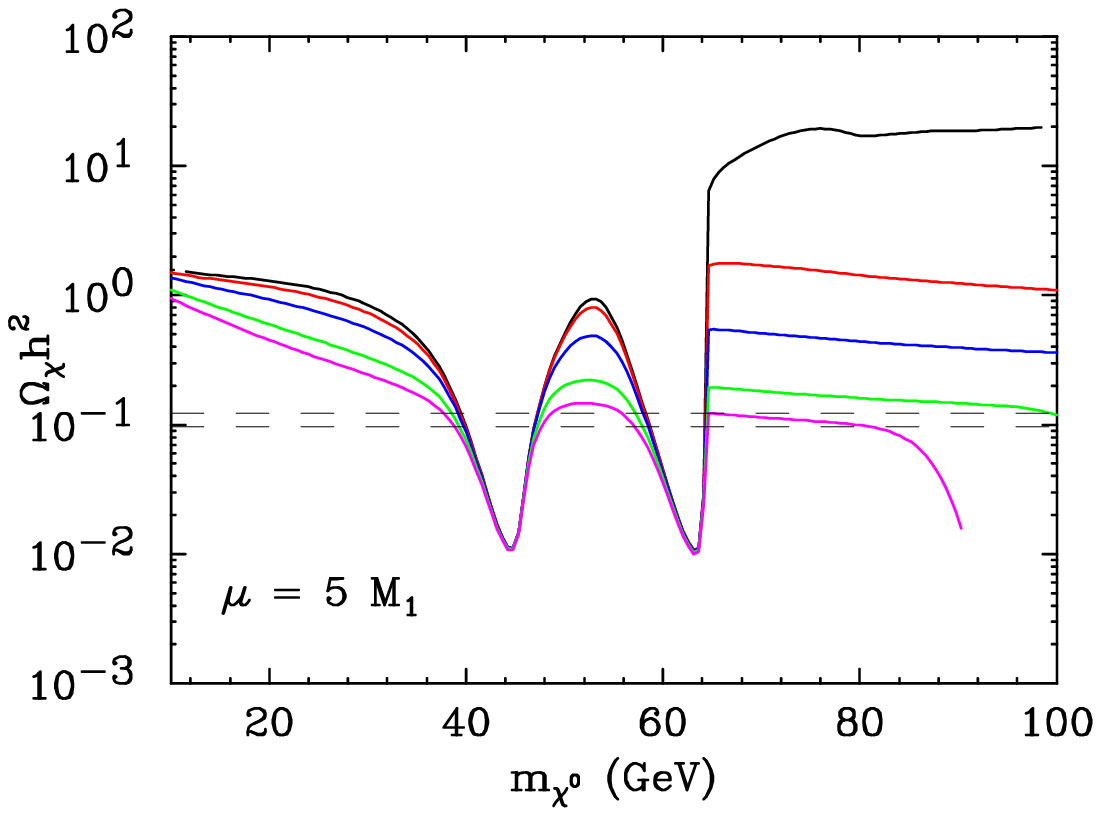}
\hspace{0.8cm}
\includegraphics[width=3.1in,angle=0]{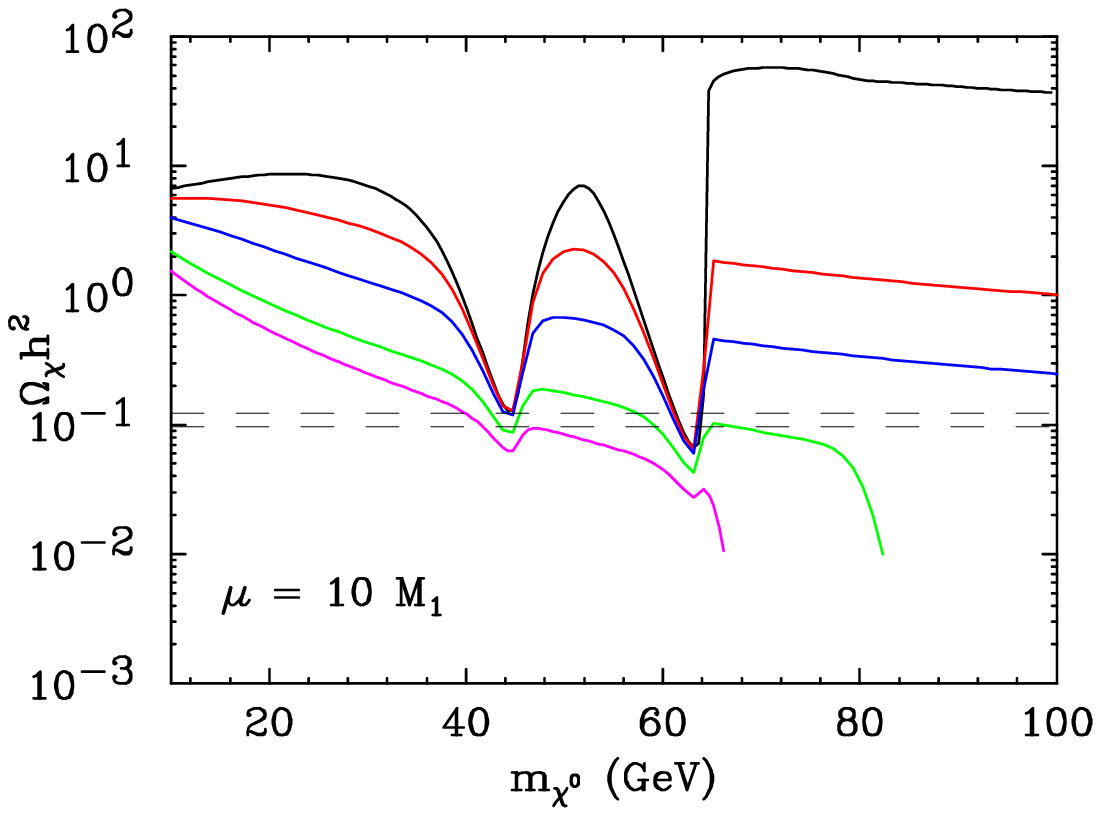}
\caption{The thermal neutralino relic abundance as a function of its mass 
  for various values of the slepton masses. The slepton masses
  neglecting mixing are (from top to bottom) 1~TeV (black), 300~GeV
  (red), 200~GeV (blue), 140~GeV (green) and 120~GeV (magenta). Other
  parameters are $m_A=m_{\tilde{q}}$=1~TeV and $\tan \beta$=10. $M_2$
  is either $2 M_1$ or the lowest value consistent with the LEP
  chargino bound, whichever is greater. The trilinear couplings $A_t$
  and $A_b$ are selected to maximize the light Higgs mass. In the left
  (right) frame, $\mu$ as set to $5 M_1$ ($10 M_1$). The horizontal
  dashed lines denote the dark matter abundance measured by
  WMAP~\cite{wmap}. The two dips correspond to the $Z$ and light Higgs
  resonances.}
\label{fig:relic1}
\end{figure}

In most supersymmetric models within the reach of trilepton searches
at the Tevatron, the lightest neutralino typically annihilates
somewhat inefficiently and thus is expected to be produced in the
early universe with a thermal abundance in excess of the measured
dark matter density. There are a number of possible exceptions to
this conclusion, however. In particular:
\begin{itemize}
  
\item[--] If the lightest neutralino is within a few GeV of the $Z$ or
  $h$ resonances ($2m_{\chi^0} \approx m_{Z,h}$), then annihilations
  through these channels can be very efficient, especially if the
  neutralino has a sizable higgsino fraction ({\it i.e.}~moderate to
  small values of $|\mu|$). For example, the cross section for
  $Z$-mediated neutralino annihilation scales simply as the square of
  the difference of the two higgsino fractions,
  $(|N_{13}|^2-|N_{14}|^2)^2$. Its effect can be seen in
  Fig.~\ref{fig:relic1}.
  
\item[--] Light sleptons, which are required in models within the
  reach of trilepton searches at the Tevatron, can also lead to
  efficient neutralino annihilation. In the extreme case, the lightest
  stau can be quasi-degenerate with the lightest neutralino, leading
  to highly efficient coannihilations. The effect of sleptons in the
  neutralino relic abundance calculation can be seen in
  Fig.~\ref{fig:relic1}. In this figure, we show the relic density as
  a function of the LSP mass, for various values (1000, 300, 200, 140
  and 120~GeV) of the slepton masses.\footnote{By ``slepton mass'' or
    $m_{\tilde{l}}$, we refer to a common mass for the selectrons,
    smuons and staus before off--diagonal terms in the mass matrices
    are accounted for. This quantity approximately corresponds to the
    selectron and smuon masses. The staus, in contrast, will depart
    somewhat from this value, $m^2_{\tilde{\tau}}\sim m^2_{\tilde{l}}
    \mp m_{\tau} (A_{\tau} - \mu \tan \beta$).}
  
\item[--] If the currently (largely) unconstrained pseudoscalar Higgs
  boson $A^0$ is light enough and its couplings are large (large $\tan
  \beta$ and/or small $|\mu|$) then it will efficiently mediate
  neutralino annihilations. When not near the $A^0$-resonance, the
  cross section to down--type fermions through pseudoscalar Higgs
  exchange is proportional to $M^2_1 \tan^2 \beta \, m^2_f/ |\mu|
  m_A^4$. This contribution is most significant in the case of a mixed
  gaugino--higgsino with a light pseudoscalar Higgs and large $\tan
  \beta$.

\end{itemize}
From Fig.~\ref{fig:relic1}, it is obvious that light neutralinos will
be overproduced in the early universe unless the sleptons are light,
the lightest neutralino's mass is within a few GeV of the $Z$ or $h$
resonances or pseudoscalar Higgs exchange provides a significant
contribution to the annihilation cross section.

\begin{figure}[t]
\includegraphics[width=2.2in,angle=0]{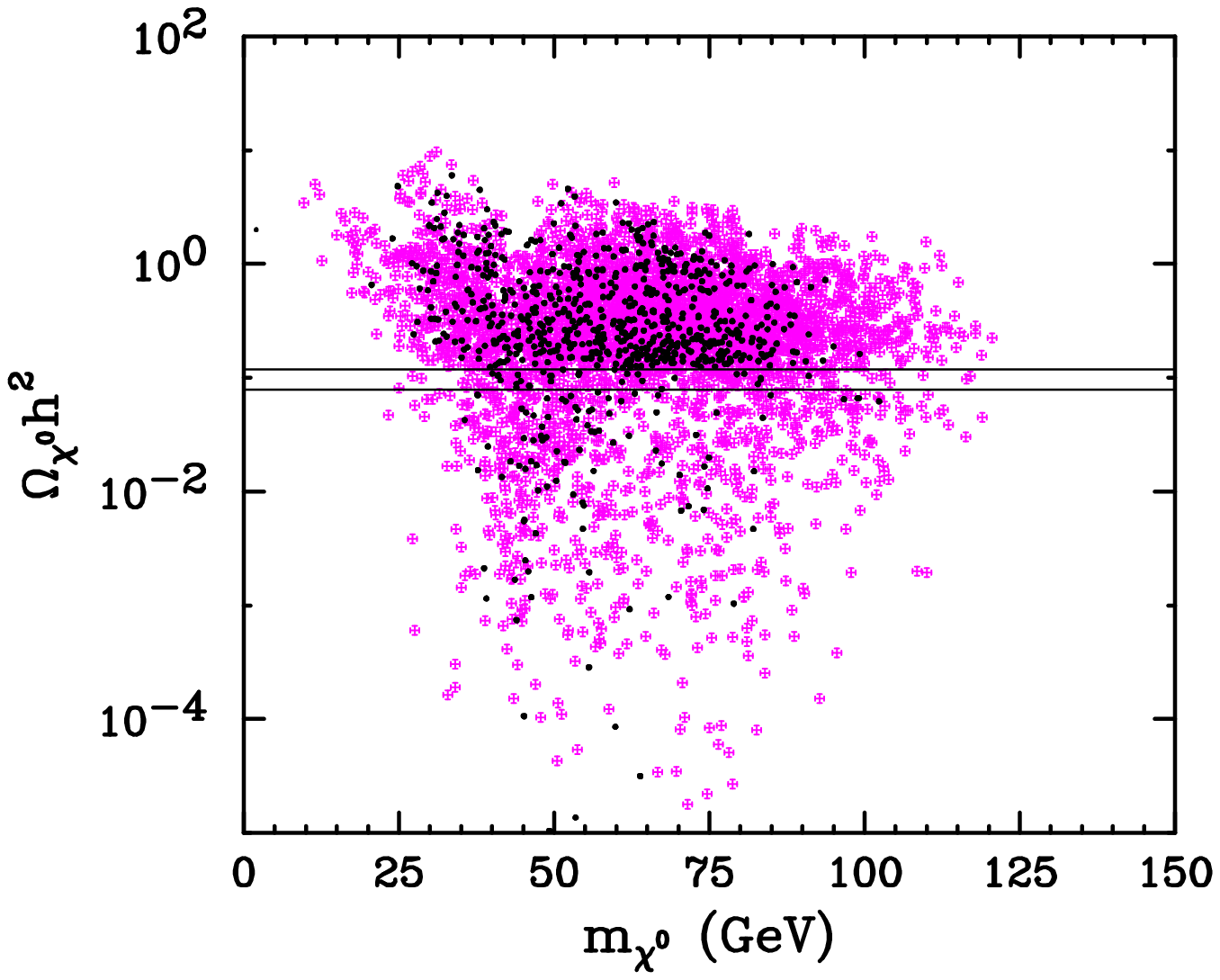} 
\includegraphics[width=2.2in,angle=0]{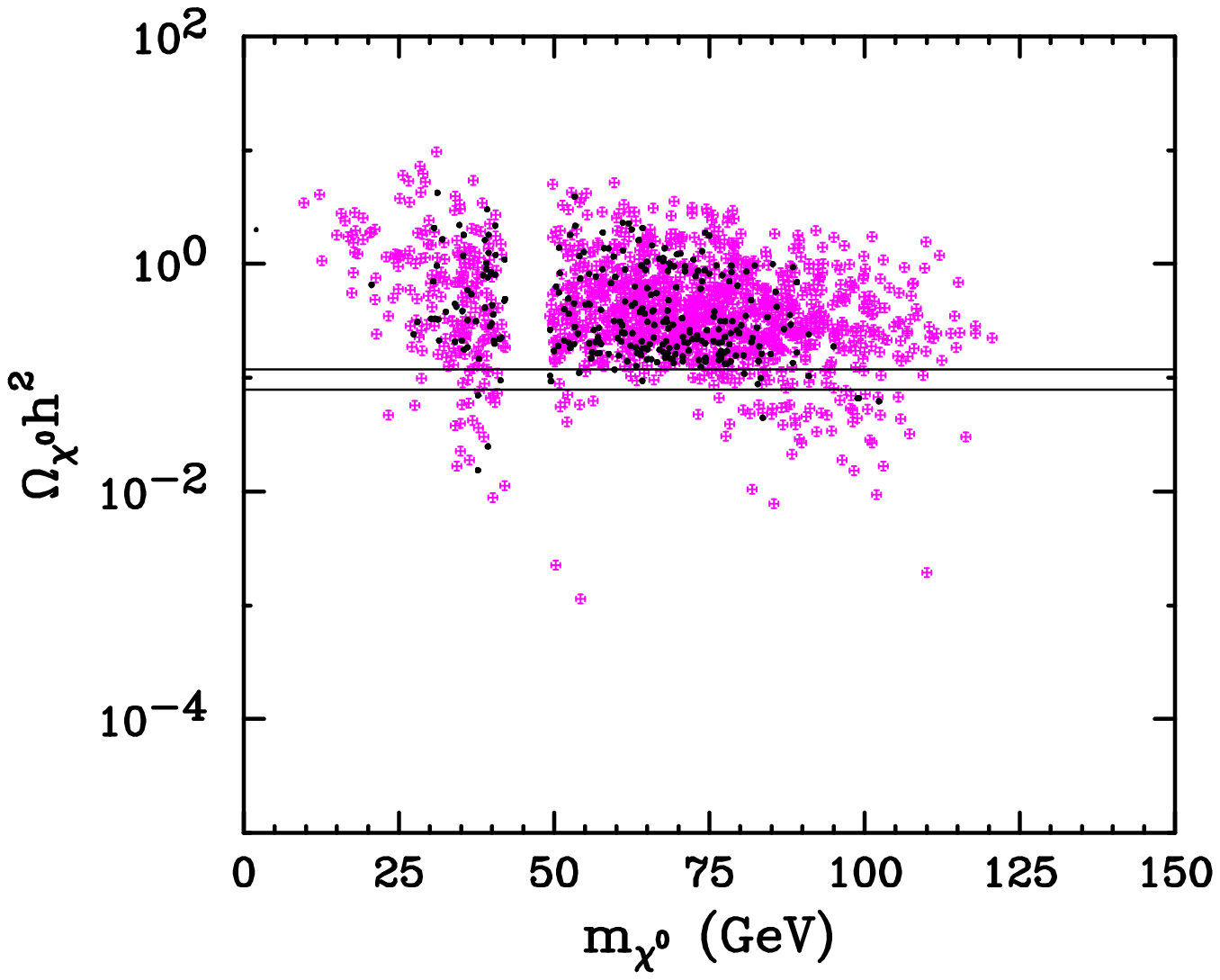}
\includegraphics[width=2.2in,angle=0]{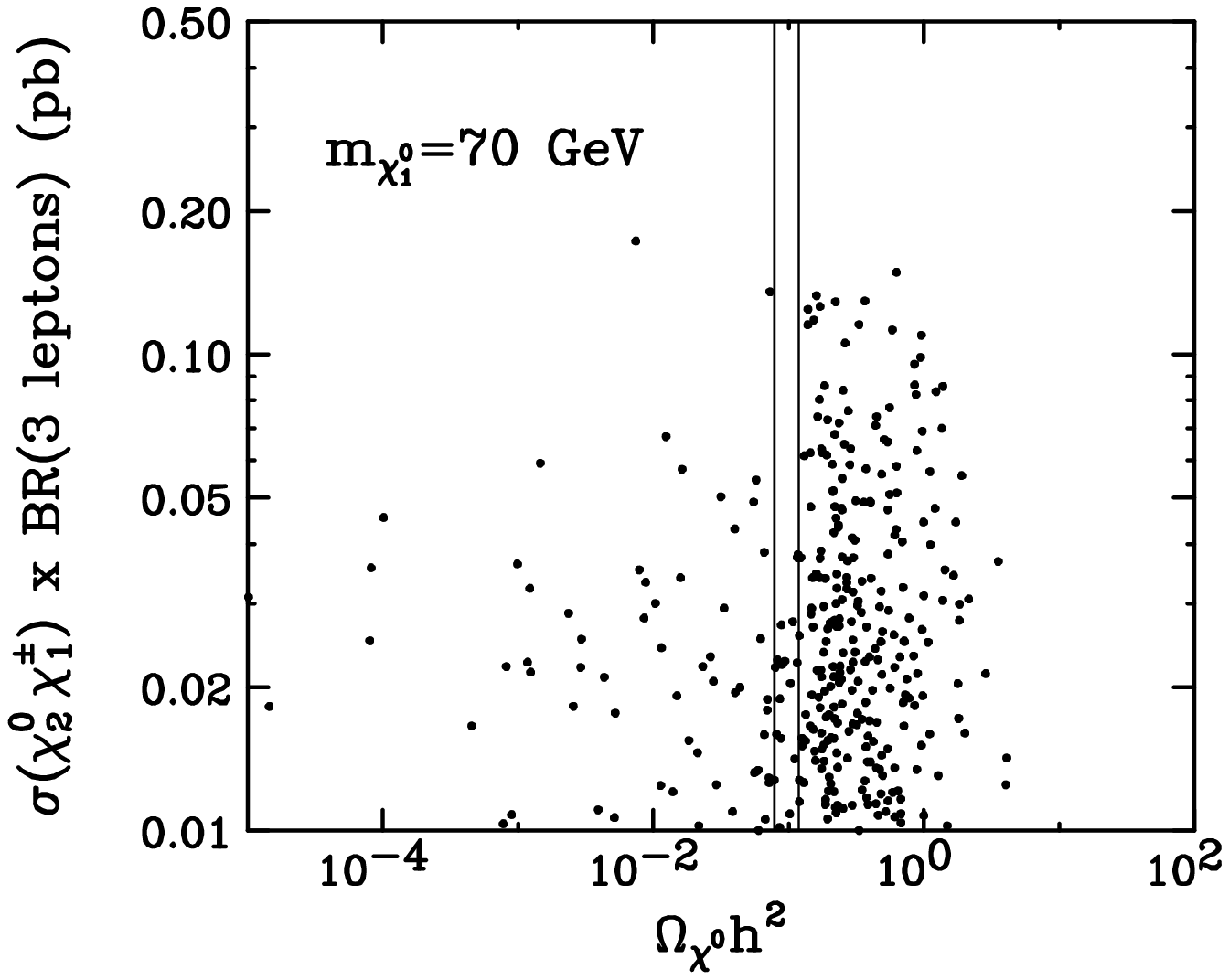}
\caption{Left panels: the thermal neutralino relic abundance as a function of 
  the neutralino mass, in models within the reach of Tevatron
  trilepton searches. Dark points denote models which have already
  been excluded by the Tevatron trilepton searches, whereas the
  lighter points are within reach with 8 fb$^{-1}$ of integrated
  luminosity. In the left frame, all of the models within the
  Tevatron reach are shown.  In the center frame, we omit models
  with efficient $Z,h$ or $A^0$-mediated dark matter annihilation (see
  text for details). In the right frame, we show the cross section times branching
  ratio for trilepton production at the tevatron as a function of the
  thermal relic abundance of neutralinos. In this frame we only show
  models with LSP masses within the range of 70$\pm$1 GeV.}
\label{fig:relic2}
\end{figure}

We demonstrate this further in Fig.~\ref{fig:relic2}, where we compare
the relic abundances found in various models within the Tevatron reach. In this parameter scan, we
vary the masses $M_1$, $M_2$, $m_{\tilde{l}}$, $m_{\tilde{q}}$,
$|\mu|$ and $m_A$ up to 1~TeV. Values of $\tan \beta$ within the range
of 1 to 60 are considered.  For simplicity, we assume the gluino mass
to be $M_3 \approx 3.7 M_2$. All models shown in Fig.~\ref{fig:relic2}
satisfy all collider constraints on the chargino, slepton, squark and
Higgs masses. The relic abundance we compute using
DarkSUSY~\cite{darksusy}.

In the left frame of Fig.~\ref{fig:relic2} we show all models found to
be within the Tevatron reach.  In the second frame we omit models with
neutralinos annihilating through either a $Z,h$ resonance or via an
$s$-channel $A^0$ diagram. To quantify this selection we remove
parameter points in which the lightest neutralino mass is within 7~GeV
of the $Z$ or $h$ resonances, or with $(\tan\beta/10)^2/[(m_A/1\,{\rm
  TeV})^4 \, (|\mu|/1\,{\rm TeV})] > 1$. These cuts overwhelmingly
remove models with low relic densities, thus demonstrating that
neutralino annihilation through $Z$ or $h$ resonances or through
$A^0$-exchange are essentially required if dark matter is to avoid
being overproduced in models within the reach of the Tevatron. Note
however, that independent of the LSP mass well always find models
which produce the correct relic density.

In the right frame of Fig.~\ref{fig:relic2} we fix the LSP mass to $70
\pm 1$~GeV and show the correlation between the trilepton cross
section times branching ratio versus the relic density. The mass of
the produced neutralino and chargino is free. The fact that the
majority of points tend towards overclosing the universe corresponds
to a bias in the entire data sample, also seen in the left two panels
of the same figure. We checked that independent of the LSP mass chosen
there is indeed no visible correlation between the relic density and
the Tevatron trilepton cross section in the MSSM.

\section{Direct Detection of Neutralinos in Tevatron Reach}
\label{sec:direct}

Experiments such as XENON~\cite{xenon}, CDMS~\cite{cdms} and many
others~\cite{others} have over the last several years placed
increasingly stringent limits on the elastic scattering cross section
of WIMPs with nuclei. The neutralino's elastic scattering cross
section with nuclei is given by
\begin{equation}
\sigma \approx \frac{4 m^2_{\chi^0} m^2_{T}}{\pi (m_{\chi^0}+m_T)^2} 
               \; \left[ Z f_p + (A-Z) f_n \right]^2,
\label{eq:sigeq}
\end{equation}
where $m_T$ is the target nuclei's mass, and $Z$ and $A$ are the
atomic number and atomic mass of the nucleus. $f_p$ and $f_n$ are the
neutralino's couplings to protons and neutrons, given by:
\begin{equation}
f_{p,n} = \sum_{q=u,d,s} f^{(p,n)}_{T_q} a_q \frac{m_{p,n}}{m_q} 
        + \frac{2}{27} \, f^{(p,n)}_{TG} 
          \sum_{q=c,b,t} a_q  \frac{m_{p,n}}{m_q},
\label{eq:feqn}
\end{equation}
where $a_q$ are the neutralino-quark couplings and $f^{(p)}_{T_u}
\approx 0.020\pm0.004$, $f^{(p)}_{T_d} \approx 0.026\pm0.005$,
$f^{(p)}_{T_s} \approx 0.118\pm0.062$, $f^{(n)}_{T_u} \approx
0.014\pm0.003$, $f^{(n)}_{T_d} \approx 0.036\pm0.008$ and
$f^{(n)}_{T_s} \approx 0.118\pm0.062$~\cite{nuc}.

The first term in the above equation corresponds to interactions with
the quarks in the target nuclei, whereas the second term denotes
interactions with the gluons in the target through a quark/squark loop
diagram. $f^{(p)}_{TG}$ is given by $1
-f^{(p)}_{T_u}-f^{(p)}_{T_d}-f^{(p)}_{T_s} \approx 0.84$, and
analogously, $f^{(n)}_{TG} \approx 0.83$.

The neutralino-quark coupling is given by~\cite{scatteraq}:
\begin{eqnarray}
a_q & = & - \frac{1}{2(m^{2}_{1i} - m^{2}_{\chi})} 
         {\rm Re} \left[ \left( X_{i} \right) 
                         \left( Y_{i} \right)^{\ast} 
                  \right] 
          - \frac{1}{2(m^{2}_{2i} - m^{2}_{\chi})} 
         {\rm Re} \left[ \left( W_{i} \right) 
                         \left( V_{i} \right)^{\ast} 
                  \right] \nonumber \\
    &   & - \frac{g_2 m_{q}}{4 m_{W} B} 
            \left[ {\rm Re} \left( \delta_1 [g_2 N_{12} - g_1 N_{11}] \right) 
                            D C \left( - \frac{1}{m^{2}_{H}} 
                                       + \frac{1}{m^{2}_{h}} \right) \right. 
                          \nonumber \\
    &   & +  {\rm Re} \left. 
                      \left( \delta_{2} [g_2 N_{12} - g_1 N_{11}] \right) 
                      \left( \frac{D^{2}}{m^{2}_{h}}
                           + \frac{C^{2}}{m^{2}_{H}} \right) \right],
\label{eq:aq}
\end{eqnarray}
where
\begin{eqnarray}
X_{i}& \equiv& \eta^{\ast}_{11} 
        \frac{g_2 m_{q}N_{1, 5-i}^{\ast}}{2 m_{W} B} - 
        \eta_{12}^{\ast} e_{i} g_1 N_{11}^{\ast}, \nonumber \\
Y_{i}& \equiv& \eta^{\ast}_{11} \left( \frac{y_{i}}{2} g_1 N_{11} + 
        g_2 T_{3i} N_{12} \right) + \eta^{\ast}_{12} 
        \frac{g_2 m_{q} N_{1, 5-i}}{2 m_{W} B}, \nonumber \\
W_{i}& \equiv& \eta_{21}^{\ast}
        \frac{g_2 m_{q}N_{1, 5-i}^{\ast}}{2 m_{W} B} -
        \eta_{22}^{\ast} e_{i} g_1 N_{11}^{\ast}, \nonumber \\
V_{i}& \equiv& \eta_{22}^{\ast} \frac{g_2 m_{q} N_{1, 5-i}}{2 m_{W} B}
        + \eta_{21}^{\ast}\left( \frac{y_{i}}{2} g_1 N_{11},
        + g_2 T_{3i} N_{12} \right).
\label{eq:xywz}
\end{eqnarray}
In these expressions, $i=1,2$ denote up and down-type quarks,
respectively. $m_{1i}, m_{2i}$ denote the squark mass eigenvalues and
$\eta$ is the matrix which diagonalizes the squark mass matrices.
$y_i$, $T_{3i}$ and $e_i$ denote hypercharge, isospin and electric
charge of the quarks. For scattering off of up-type quarks $\delta_1 =
N_{13}, \delta_2 = N_{14}, B = \sin \beta, C = \sin \alpha, D = \cos
\alpha$, whereas for down-type quarks $\delta_1 = N_{14}, \delta_2 =
-N_{13}, B = \cos \beta, C = \cos \alpha, D = -\sin \alpha$. $\alpha$
is the mixing angle in the Higgs sector.\bigskip

\begin{figure}[t]
\begin{center}
\includegraphics[width=3.3in,angle=0]{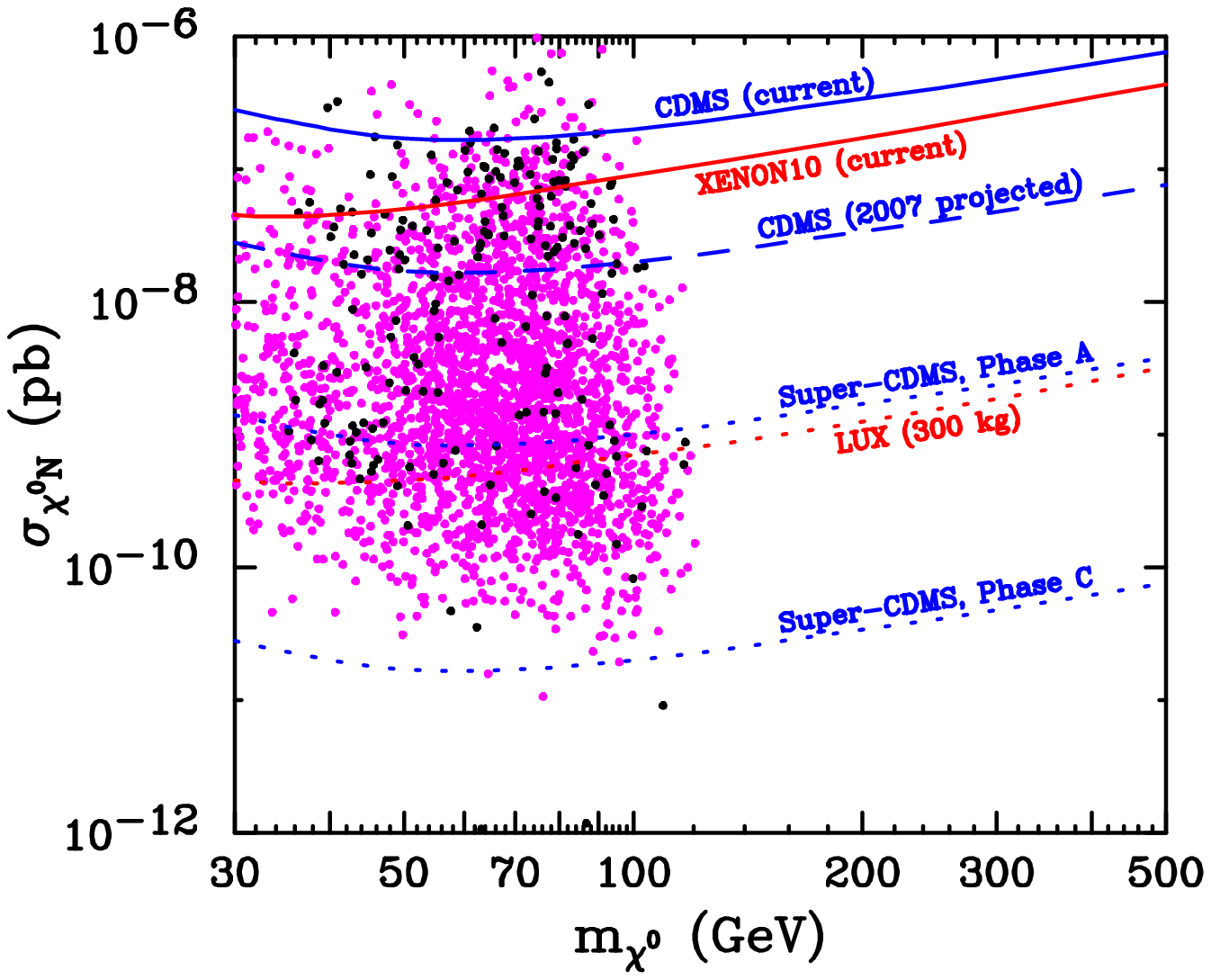}
\includegraphics[width=3.3in,angle=0]{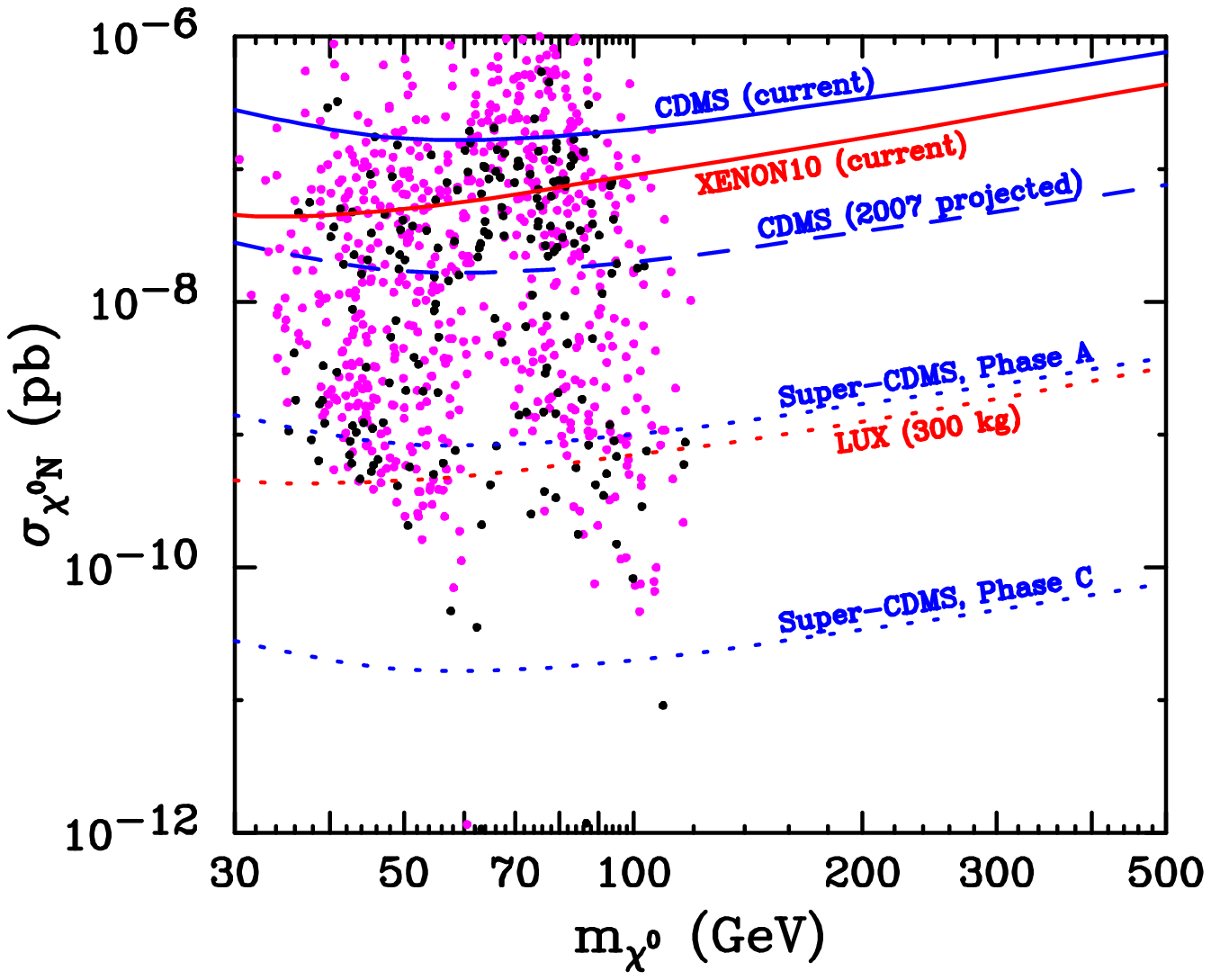}
\end{center}
\caption{The neutralino's elastic scattering cross section with nucleons, as a function of its mass, in model which are within the reach of trilepton searches at the Tevatron with 8 fb$^{-1}$ luminosity. In both frames, the dark points represent models which are predicted to generate a thermal density of dark matter within the range measured by WMAP~\cite{wmap}. In the left (right) frame, the lighter points represent models which predict a larger (smaller) dark matter density than is measured.}
\label{fig:direct}
\end{figure}

In Fig.~\ref{fig:direct}, we plot the neutralino-nucleon elastic
scattering cross section found in those models which are within the 8
fb$^{-1}$ reach of the Tevatron's trilepton search. In each frame, the
dark points correspond to models which predict a thermal abundance of
neutralino dark matter within the range measured by WMAP~\cite{wmap}.
The lighter points represent models with too much (left) or too little
(right) dark matter relative to the measured abundance. As expected
from the different composition patterns of the neutralinos, we find a
very large range of cross sections, varying from about $10^{-6}$ to
$10^{-11}$~pb. Models which predict an abundance of dark matter below
the measured value tend to have somewhat larger elastic scattering
cross sections with nuclei. This is due to the coupling to heavy MSSM
Higgs bosons which can both mediate neutralino annihilation and
elastic scattering processes.  Most models with the observed relic
density fall in the upper portion of the elastic cross section range.
In particular, the majority of them are within the reach of CDMS's
current run (labelled CDMS 2007). This is not the case for a typical
scan over the entire MSSM parameter space, which consists mostly of
models beyond the Tevatron's reach (see, for example,
Ref.~\cite{cdms_sensitivity}).

The reasons for the favorable direct detection prospects among models
within the reach of the Tevatron are somewhat subtle. In the case of a
neutralino annihilating in the early universe primarily through a $Z$
or $h$ resonance, little can be said regarding the prospects for
direct detection. Furthermore, in models which annihilate largely
through slepton exchange in the early universe (or through
coannihilations with sleptons), the elastic scattering cross section
is likely to be suppressed. In many of the models within the reach of
the Tevatron, however, the neutralino annihilation cross section is
dominated by pseudoscalar Higgs exchange. In these models, which
feature moderate to large values of $\tan \beta$ and somewhat light
pseudoscalar Higgs masses, the elastic scattering cross section is
typically dominated by the exchange of the heavy scalar Higgs, $H$,
with strange and bottom quarks, leading to a neutralino-nucleon cross
section of:
\begin{equation}
\sigma_{\chi N} \sim \frac{g^2_1 g^2_2}{4\pi} \;
                    \frac{1}{m^2_W \cos^2 \beta} \;
                    \frac{m^4_N}{m^4_H} \; 
                    |N_{11}|^2 |N_{13}|^2 \; 
                    \left(f_{T_s}+\frac{2}{27}f_{TG} \right)^2.
\label{eq:case1}
\end{equation}
The similarity $m_H \sim m_A$ is of course not an artefact of a
SUSY--breaking assumption, but a generic feature of the
two--Higgs--doublet model.  Because in a large fraction of these
models the combination of $\tan^2 \beta /m_A^4 |\mu|$ is large in
order to generate an acceptable relic abundance, the direct detection
rates also have a tendency to be larger compared to those found in a
more general sample of supersymmetric models.

\section{Outlook}
\label{sec:conclusion}

In this article, we have studied the cosmological implications of
supersymmetric models within the reach of searches for associated
neutralino-chargino production at Run~II of the Tevatron. We have
analyzed how results from this Tevatron search channel might impact
the prospects for direct searches for neutralino dark matter. Although
there is not a particularly direct or obvious connection between these
two experimental programs, it is important to consider how to exploit
the interplay between collider and astrophysical searches for
supersymmetry.\bigskip

Supersymmetric models whose rate of trilepton events from associated
chargino-neutralino production is within the reach of the Tevatron
have some rather peculiar features. In particular, they contain light
neutralinos which either annihilate through a $Z$ or $h$ resonance,
through pseudoscalar Higgs exchange, or via very light sleptons.
Therefore, for models in which the lightest neutralino is not within a
few GeV of $m_Z/2$ or $m_h/2$, the heavy Higgs bosons $A, H$ tend to
be light and values of $\tan \beta$ are typically moderate to large.
These neutralinos will also show a non--negligible higgsino fraction.
These features lead to larger than average elastic scattering cross
sections with nuclei (which is dominated by $H$ exchange), and high
rates in underground direct dark matter experiments.  This means that
if the Tevatron detects trilepton events from associated
neutralino--chargino production, the near future prospects for the
direct detection of neutralino dark matter are expected to be
promising.\bigskip

The absense of more distinct parameter correlations means that the
collider and the cosmological analyses of the MSSM neutralino and
chargino sector probe different properties of the supersymmetric spectrum. This is different from the case of, for example, gravity--mediated SUSY breaking.  Looking at the TeV--scale MSSM this
implies that the information gained in dark matter searches is
largely orthogonal to the information which could be obtained from collider searches. Only by combining many sets of information from many different experimental channels will it become possible to construct with confidence a consistent picture of the TeV--scale Lagrangian~\cite{sfitter}. 

\acknowledgments{DH is supported by the US Department of Energy and by
  NASA grant NAG5-10842. Fermilab is operated by Fermi Research
  Alliance, LLC under Contract No.~DE-AC02-07CH11359 with the United
  States Department of Energy. AV thanks the Agence National de la
  Recherche for providing financial support.}

\end{document}